\documentclass[prd,superscriptaddress,twocolumn,preprintnumbers,%
  amsmath,amssymb]{revtex4-2}
\usepackage[utf8]{inputenc}

\usepackage{amsmath,bm}
\usepackage{graphicx,csvsimple,booktabs,epsfig}

\date{\today}

\usepackage[breaklinks,colorlinks=true]{hyperref}
\usepackage{mathtools}
\usepackage{amsthm,amssymb}
\usepackage{lipsum}
\usepackage{calrsfs}
\usepackage{color}
\allowdisplaybreaks
\usepackage[dvipsnames]{xcolor}
\newcommand\myshade{85}
\colorlet{mylinkcolor}{BrickRed}
\colorlet{mycitecolor}{NavyBlue}
\colorlet{myurlcolor}{Aquamarine}
\hypersetup{
  linkcolor  = mylinkcolor!\myshade!black,
  citecolor  = mycitecolor!\myshade!black,
  urlcolor   = myurlcolor!\myshade!black,
  colorlinks = true,
}

\newcommand{\Mod}[1]{\ (\mathrm{mod}\ #1)}

\begin{document}
\title{Quantifying high-order interdependencies on individual patterns via\\ the local O-information:
theory and applications to music analysis}

\author{Tomas Scagliarini}
\affiliation{Dipartimento Interateneo di Fisica, Universitá degli Studi Aldo Moro, Bari and INFN, Italy}
\author{Daniele Marinazzo}
\affiliation{Department of Data Analysis, Ghent University, Belgium}
\author{Yike Guo}
\affiliation{Department of Computer Science, Hong Kong Baptist University, Hong Kong}
\affiliation{Data Science Institute, Imperial College London, United Kingdom}
\author{Sebastiano Stramaglia}
\affiliation{Dipartimento Interateneo di Fisica, Universitá degli Studi Aldo Moro, Bari and INFN, Italy}
\affiliation{Center of Innovative Technologies for Signal Detection and Processing (TIRES), Universitá degli Studi Aldo Moro, Italy}
\author{Fernando E. Rosas}
\affiliation{Data Science Institute, Imperial College London, United Kingdom}
\affiliation{Centre for Psychedelic Research, Department of Brain Science, Imperial College London, United Kingdom}
\affiliation{Centre for Complexity Science, Imperial College London, United Kingdom}

\begin{abstract}

High-order, beyond-pairwise interdependencies are at the core of biological, economic, and social complex systems, and their adequate analysis is paramount to understand, engineer, and control such systems. 
This paper presents a framework to measure high-order interdependence that disentangles their effect on each individual pattern exhibited by a multivariate system. 
The approach is centred on the \emph{local O-information}, a new measure that assesses the balance between synergistic and redundant interdependencies at each pattern. 
To illustrate the potential of this framework, we present a detailed analysis of music scores from J.S. Bach, which reveals how high-order interdependence is deeply connected with highly non-trivial aspects of the musical discourse. Our results place the local O-information as a promising tool of wide applicability, which opens new perspectives for analysing high-order relationships in the patterns exhibited by complex systems.

\end{abstract}
\maketitle

\section{Introduction}
The analysis of interdependence is crucial for understanding the staggering complexity of structures and behaviours manifested in biological, economic and social systems. 
The unprecedented amount of data available for scientific scrutiny provides unique opportunities to deepen our understanding of 
multivariate co-evolving complex systems, including the orchestrated activity of multiple brain areas,
the interactions between different genes, and 
the relationship between various econometric indices. Importantly, what allows these systems to be more than the sum of their parts is not to be found in the material nature of their parts, but in the fine structure of their interdependencies~\cite{crutchfield_calculi_1994}.

Information theory provides an ideal framework to study interdependencies in multivariate system, which establishes the notion of  \emph{information} as a common currency under which diverse systems can be measured and compared~\cite{crutchfield_regularities_2003}.  
A particularly promising approach for analysing the structure of interdependencies is the partial information decomposition (PID), which distinguishes different `modes' of information that multiple predictors convey about a target variable~\cite{williams2010nonnegative,griffith_quantifying_2014,lizier_information_2018}. Two paradigmatic examples of such modes are synergy and redundancy~\cite{gat_synergy_1999,schneidman_synergy_2003,timme_synergy_2014,rosas_understanding_2016,wibral_partial_2017}: redundancy 
corresponds to information which can be retrieved independently from more than one source, while synergy correspond to statistical relationships that exist in the whole but cannot be seen in the parts --- this being rooted in the elementary fact that variables can be pairwise independent while being globally correlated.

Despite continuous efforts to develop PID, the precise way in which synergies and redundancies should be calculated is still being revised~\cite{harder_bivariate_2013,griffith_intersection_2014,barrett_exploration_2015,olbrich_information_2015,ince_partial_2017,finn_pointwise_2018,james_unique_2018,james_unique_2019,ay_information_2019,rosas_operational_2020,finn_generalised_2020,gutknecht_bits_2021,schick-poland_partial_2021,makkeh_introducing_2021}
. 
One way to circumvent this challenge is to avoid computing the full decomposition, and study mixtures of PID modes that can be captured by linear combinations of Shannon measures. 
One such measure is the O-information~\cite{rosas_quantifying_2019}, which has been shown to effectively capture the overall balance between redundant and synergistic modes. 
The effectiveness of the O-information in practical analyses has been verified by recent applications on populations of spiking neurons~\cite{stramaglia_quantifying_2021}, and the relationship between neural patterns and ageing~\cite{gatica_high-order_2021}.

An important limitation of the O-information is that it characterises a multivariate system with a single number, which summarises to the aggregated effect of various patterns. 
Building on the rich literature of pointwise information measures~\cite{bossomaier_transfer_2016,lizier_local_2008,stramaglia_local_2021}, in this paper we introduce the \emph{local O-information}, which evaluates each pattern separately --- such that its ensemble average recovers the O-information.
More specifically, the local O-information constitutes an overall measure that characterise the high-order interdependencies between the parts of a multivariate system at each possible pattern of activity. 
Put differently, the local O-information evaluates the `statistical quality' of each pattern, providing a signed scalar that assesses the balance between redundancies an synergies at each individual pattern.





This paper presents the theory behind the local O-information, and then illustrates its rich capabilities by analysing the scores of the chorales of J.S. Bach. Our results show how the local O-information is capable of revealing subtle musical relationships, including properties of different intervals, chord dispositions, harmonic depth, and the relationship between music and text. 
Thanks to its ability to uncover such highly non-trivial relationships, the local O-information is a valuable addition to the toolkit of data analysts interested in the study of complex systems.

The rest of the paper is organised as follows. Section II provides background information about the O-information, introduces the new local O-information, and then illustrates its basic properties on small spin systems. Then, Section III presents a detailed analysis of the local O-information on the chorales of J.S Bach, and finally Section VI summarises our main conclusions.

\section{A local measure of information quality}

Let us consider a scientist who is interested in studying a given complex system, whose state can be appropriately described by the vector $\bm{X}^n = (X_1, \dots, X_n)$. 
We focus on scenarios where the scientist has enough data to build a reliable statistical description of the statistics of $\bm{X}^n$, which is denoted by $p(\bm{X}^n)$. A question of interest is how to leverage the statistics encoded in $p$ in order to deepen our understanding of the structure of interdependencies that characterize $\bm{X}^n$. Such understanding can lead either to the building of statistical markers to classify different systems or different states of the same system, or to establish parallels between seemingly heterogeneous systems based on the similarity of their relational structure.

Through this section, random variables are denoted by capital letters (e.g. $X,Y$) and their realisations by lower case letters (e.g. $x,y$). Random vectors and their realisations are denoted by capital and lower case boldface letters, respectively.

\subsection{O-Information}


Shannon's mutual information is a popular metric of interdependency, which overcomes the limitations of correlation metrics such as Pearson's in that it captures both linear and non-linear relationships and being applicable to ordinal data. However, the mutual information can only asses the relationships between two (sets of) variables, being unable to fully explore the rich interplay that can take place within triple or higher-order interactions.

Two multivariate extensions of the mutual information are the \textit{Total Correlation} (TC)~\cite{watanabe_information_1960} and the \textit{Dual Total Correlation} (DTC)~\cite{te_sun_nonnegative_1978}, which are defined as
\begin{align*}
\text{TC}(\bm{X}^n) &:= \sum_{i=1}^n H(X_i) - H(\bm X^n) , \\
\text{DTC}(\bm{X}^n) &:= H(\bm{X}^n) - \sum_{i=0}^N H( X_i \mid \bm{X}^n_{-i}) .
\end{align*}
Above, $H(X_i) =-\sum_{x_i} p(x_i) \log p(x_i)$ corresponds to the Shannon entropy, $H(X_i|X_j) = H(X_i,X_j) - H(X_j) $ is the conditional Shannon entropy, 
and $\bm{X}^n_{-i}$ is the vector of all variables except $X_i$ (i.e., $\bm{X}^n_{-i}=\left( X_1, \ldots, X_{i-1}, X_{i+1}, \ldots, X_n \right)$); hence, the term $H( X_i \mid \bm{X}^n_{-i})$ quantifies how $X_i$ is independent from the other $n-1$ variables. 
As the mutual information, both $\text{TC}$ and $\text{DTC}$ are non-negative quantities which are zero if and only if all variables $X_0, \dots, X_N$ are jointly statistically independent --- i.e. if $p(\bm X^n) = \prod_{i=1}^n p(X_i)$.

Despite the similarities between $\text{TC}$ and $\text{DTC}$, these two metrics provide distinct but complementary views on the strength of the interdependencies in a multivariate system. On the one hand, the $\text{TC}$ accounts for \emph{collective constraints}, which refer to regions of the phase space that the system explore less; while on the other hand, the $\text{DTC}$ measures the degree of \emph{shared randomness} between the variables, i.e. the amount of information that can be collected in one variable that also refers to the activity of another~\cite{rosas_quantifying_2019}. 
An attractive way to exploit these complementary views is by considering their difference,
\begin{align}
  \Omega_n(\bm{X}^n) &= \text{TC}(\bm{X}^n) - \text{DTC}(\bm{X}^n)~. 
  \label{eq:Oinfo}
\end{align}
which is known as the \emph{O-information}~\cite{rosas_quantifying_2019}. The O-information can be seen as a revision of the measure of neural complexity proposed by Tononi, Sporns and Edelman in~\cite{tononi_measure_1994}, which provides a mathematical construction that is closer to their original desiderata~\footnote{For a comparison between the original measure proposed in Ref.~\cite{tononi_measure_1994} and the O-information, please see Ref.~\cite{rosas_quantifying_2019}.}. In effect, the O-information is a signed metric that captures the balance between high- and low-order statistical constraints~\footnote{While low-order constraints impose strong restrictions on the system and allow little amount of shared information between variables, high-order constraints impose collective restrictions that enable large amounts of shared randomness.}. 
By construction, $\Omega (\bm{X}^n) < 0$ implies a predominance of high-order constraints within the system $\bm{X}^n$, a condition that is usually referred to as \textit{statistical synergy}. Conversely, $\Omega (\bm{X}^n) > 0$ implies that the system $\bm{X}^n$ is dominated by low-order constraints, which imply \textit{redundancy} of information. This nomenclature is further supported by the following key properties:
\begin{itemize}
    \item[(1)] It captures genuine high-order effects, at it is zero for systems with only pairwise interdependencies: if the joint distribution of $\bm X^{n-1}$ (for $n$ odd) can be factorised as $p_{\bm X^n}(\boldsymbol x^n) = \prod_{k=1}^{n/2}p_{X_{2k},X_{2k+1}}(x_{2k},x_{2k+1})$, then $\Omega(\bm X^n)=0$.
    \item[(2)] The O-information is maximised by redundant distributions where the same information is copied in multiple variables, and is minimised by synergistic (``\mbox{\texttt{xor}-like}'') distributions: e.g. for binary variables, $\Omega$ is maximised by the ``n-bit copy'' where $X_1$ is a Bernoulli r.v. with parameter $p=1/2$ and $X_0=\dots=X_{N-1}$, and is minimised when $X_0,\dots,X_{N-2}$ are i.i.d. fair coins and $X_{N-1}=\sum_{j=0}^{N-2} X_j \Mod{2}$.
    \item[(3)] The O-information characterises the dominant tendency, being additive over non-interactive subsystems: if the system can be factorised as $p_{\bf X^N}(\boldsymbol x^N) = p_{X_1,\dots,X_{m}}(x_1,\dots, x_m) \times p_{X_{m+1},\dots,X_N}(x_{m+1},\dots, x_N)$, then $\Omega(\bm X^n)= \Omega(X_1,\dots,X_{m-1}) + \Omega(X_m,\dots,X_n)$.
\end{itemize}
For more details related to the O-information, we refer the reader to Ref.~\cite{rosas_quantifying_2019}.


\subsection{Local O-Information}

Building on the properties of the O-information reviewed in the previous section, one can design a 
measure that can capture these effects on a state-by-state basis. In effect, the O-information provides a single scalar that characterises the interdependencies of a system \emph{on average}. However, in many systems of interest this average represents a mid-point between highly heterogeneous states, which could potentially be of limited value to understand the role of individual patterns. In this section we introduce a \emph{pointwise} measure that allows to calculate the O-information on individual states. 

As a first step, let us introduce a \emph{local total correlation} and \emph{local dual total correlation} which are given by
\begin{align}
\text{tc}(\bm{x}^n) :=& \sum_{j=1}^nh(x_j) - h(\bm{x}^n),\\
\text{dtc}(\bm{x}^n) :=& \:h(\bm{x}^n) - \sum_{j=1}^n h(x_j|\bm{x}^n_{-j}),
\end{align}
where $h(\bm{x}^n) = -\log p(\bm{x}^n)$ is the information content of the state $\bm{x}^n$ \cite{bossomaier_transfer_2016}. These quantities capture how the strength of the multivariate interdependencies 
vary 
with the state of the system, providing a generalisation to the pointwise mutual information introduced in Ref \cite{fano1968transmission}. 

Using these definitions, we can define the \emph{local O-information} as follows:
\begin{align}
  \omega(\bm{x}^n) 
  :=& \;\text{tc}(\bm{x}^n) - \text{dtc}(\bm{x}^n) \\
  =& \;(n-2)h(\bm{x}^n) + \sum_{j=1}^n\Big(h(x_j) - h(\bm{x}_{-j}^n)\Big).
  \label{eq:localO}
\end{align}
In contrast to $\Omega(\bm X^n)$, which provides a single value for the random variable, $\omega(\bm x^n)$ assigns a number of each possible state $\bm x^n$. In particular, the local O-information has the following useful properties:
\begin{itemize}
    \item[$\bullet$] $\Omega(\bm X^n) = \mathbb{E}\{\omega(\bm x)\}$.
    \item[$\bullet$]  $\inf_{\bm x}\omega(\bm x) \leq \Omega(\bm X^n) \leq \sup_{\bm x}\omega(\bm x)$.
\end{itemize}
Generally speaking, the local O-information provides more information than its global counterpart as, technically, $\omega(\bm X^n)$ is a random variable whose mean value is $\Omega(\bm X^n)$. Therefore, the whole range of values of $\omega(\bm X^n)$ can naturally provide a more fine-grained description of the system than its mere first moment.

Additionally, note that Lemma~3 of Ref.~\cite{rosas_quantifying_2019} provides upper and lower bounds for $\Omega(\bf{X})$ when the variables take values in a finite alphabet; in particular, if $X_j \in\mathcal{X}$ for all $j$ then
\begin{equation}\label{eq:bounds}
    -(n-2) \log |\mathcal{X}|
    \leq
    \Omega(\bm{X}^n) 
    \leq (n-2)\log |\mathcal{X}|~.
\end{equation}
Note that these bounds don't apply to the local O-information --- they apply to its average value, but extreme values can be larger than it. Nonetheless, the quantity $(n-2)\log |\mathcal{X}|$ establishes a natural bound which that, when surpassed, values of $\omega$ can be considered to be particularly large. This provides a useful rule of thumb to interpret the ranges of values obtained by evaluations of $\omega$.

\subsection{Proof of concept}

An useful way of employing $\omega$ is to classify states among different \emph{types}. In particular, building on the properties of the O-information, we say that a state $\bm x^n$ for which $\omega(\bm{x}^n)>0$ is redundancy-dominated, while if $\omega(\bm{x}^n)<0$ we say the state is synergy-dominated. In this section we illustrate this capability of the local O-information by using it to analyse a small Ising system. 

Let's consider three coupled spins denoted by $(S_1,S_2,S_3)$, whose joint probability distributions follows a Boltzmann-Gibbs distribution of the form:
\begin{equation}
    p(s_1,s_2,s_3) = \frac{e^{J(s_1 s_2 +s_1 s_3+s_2 s_3)}}{Z},
    \label{hamiltonian}
\end{equation}
with $Z= \sum_{s1,s2,s3} e^{J(s_1 s_2 +s_1 s_3+s_2 s_3)}$ is a normalisation factor. For positive values of $J$, the configurations with all the spins in agreement (i.e. $\uparrow \uparrow \uparrow$ and $\downarrow \downarrow \downarrow$) satisfy all bonds, and hence the system may be seen as a small ferromagnet. In contrast, for negative $J$ the system is said to be ``frustrated'' as there is not a configuration satisfying all bonds simultaneously. 

\begin{figure}[t!]
    \centering
    \includegraphics[width=0.45\textwidth]{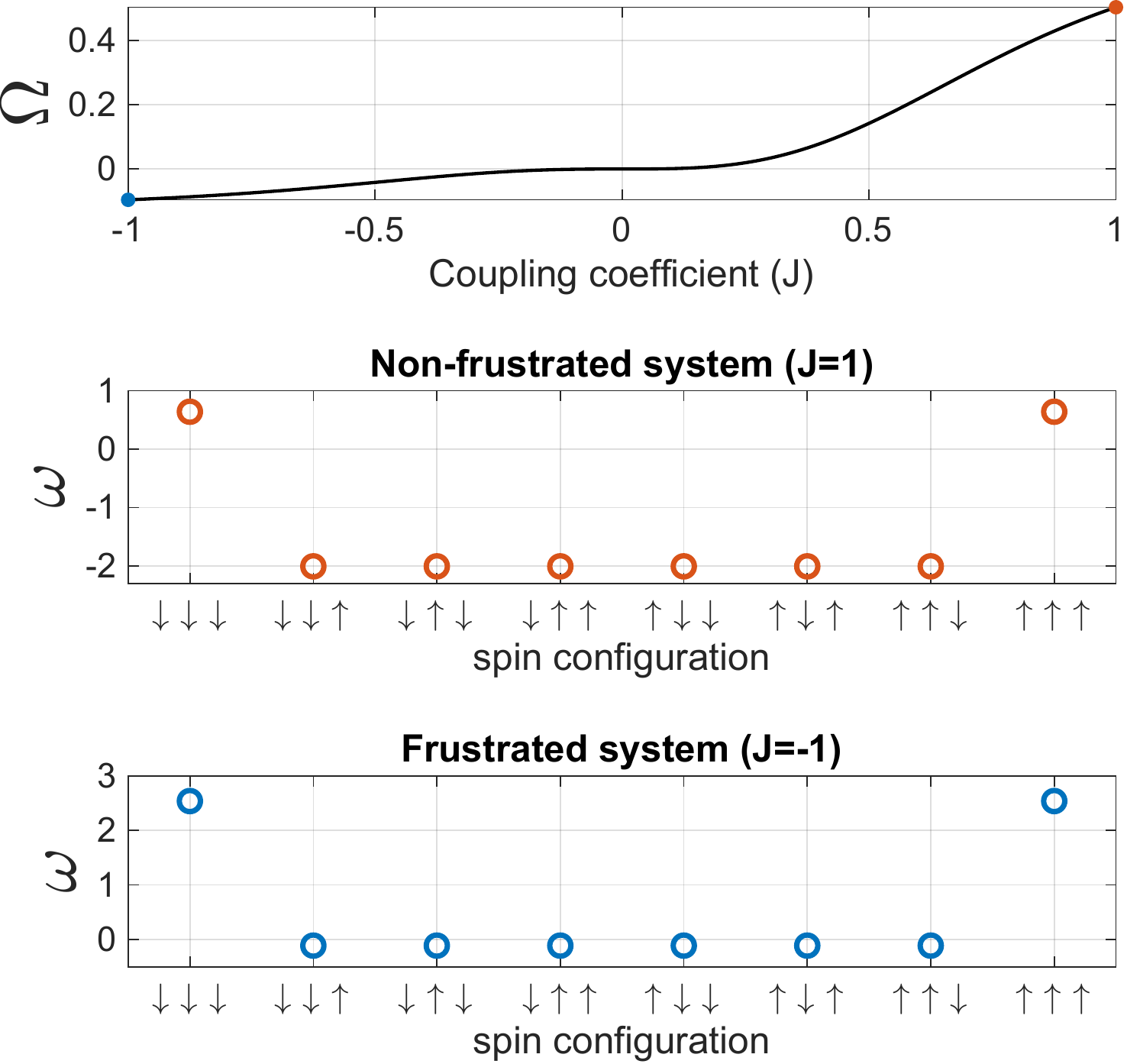}
    \caption{\textbf{Concerning the Ising toy model of three spins, $\Omega$ is plotted versus the coupling $J$ (top panel). The values of $\omega$ for the $8$ possible configurations of the three spins are shown for the unfrustrated case when $J=1$ (middle panel) and for the frustrated case, when $J=-1$ (bottom panel).}  }
    \label{fig:toy1}
\end{figure}

By analysing Eq.~\eqref{hamiltonian} via the O-information, one finds that ferromagnetic behaviour are redundancy-dominated while frustrated systems are synergy-dominated (see Figure \ref{fig:toy1}). Interestingly, the local O-information shows that configurations of spin agreement are redundancy-dominated states, whilst the six configurations with disagreement are synergy-dominated --- this for both positive and negative values of $J$. This let us conclude that what makes the system redundancy- or synergy-dominated for different values of $J$ is the different frequency with which either redundancy- or synergy-dominated configurations are visited.

This toy example shows how the local O-information can reveal different qualities of various states --- in this case, either states with agreement or disagreement. Furthermore, this example also reveals an intriguing connection between synergy and frustration in spin models, which will be further investigated in a future publication.

\section{Case study: high-order relationships in Bach's chorales}

To illustrate the usefulness of the local O-information for practical data analysis, this section
presents a study of the multivariate statistics of musical scores from the
Baroque period. 
Ref.~\cite{rosas_quantifying_2019} provided an analysis of music scores based on the global O-information; however, such analyses could not provide information about individual chords, and hence could not explore further musical aspects related to harmony and tonality. 
Here we show how the local O-information can greatly expand this type of analyses, revealing subtle aspects of the musical discourse that are reflected in the high-order interactions.

In the following, Section~\ref{sec:pipeline} describes the procedure
to obtain and analyse the data, and Section~\ref{sec:results} discusses our main findings.

\subsection{Processing pipeline}~\label{sec:pipeline}
\vspace{-1cm}

\subsubsection{Data}

Our analysis focuses on the chorales for four voices (soprano, alto, 
tenor, and bass) written by Johann Sebastian Bach (1685-1750). These works are 
characterised by an elaborate counterpoint between the melodic lines that leads 
to rich harmonic progressions, which in turn results into a broad range of chords 
displayed along the repertoire. 
An additional point of interest of these pieces is that, as typical in the 
Baroque period (approx. 1600--1750), they display a balance in the interest and richness 
of each of the four voices. This contrasts with the subsequent Classic (1730--1820) 
and Romantic (1780--1910) periods, where higher voices tend to take the lead while 
the lower voices provide mere support.

\begin{figure}
    \centering
    \includegraphics[width=0.5\textwidth]{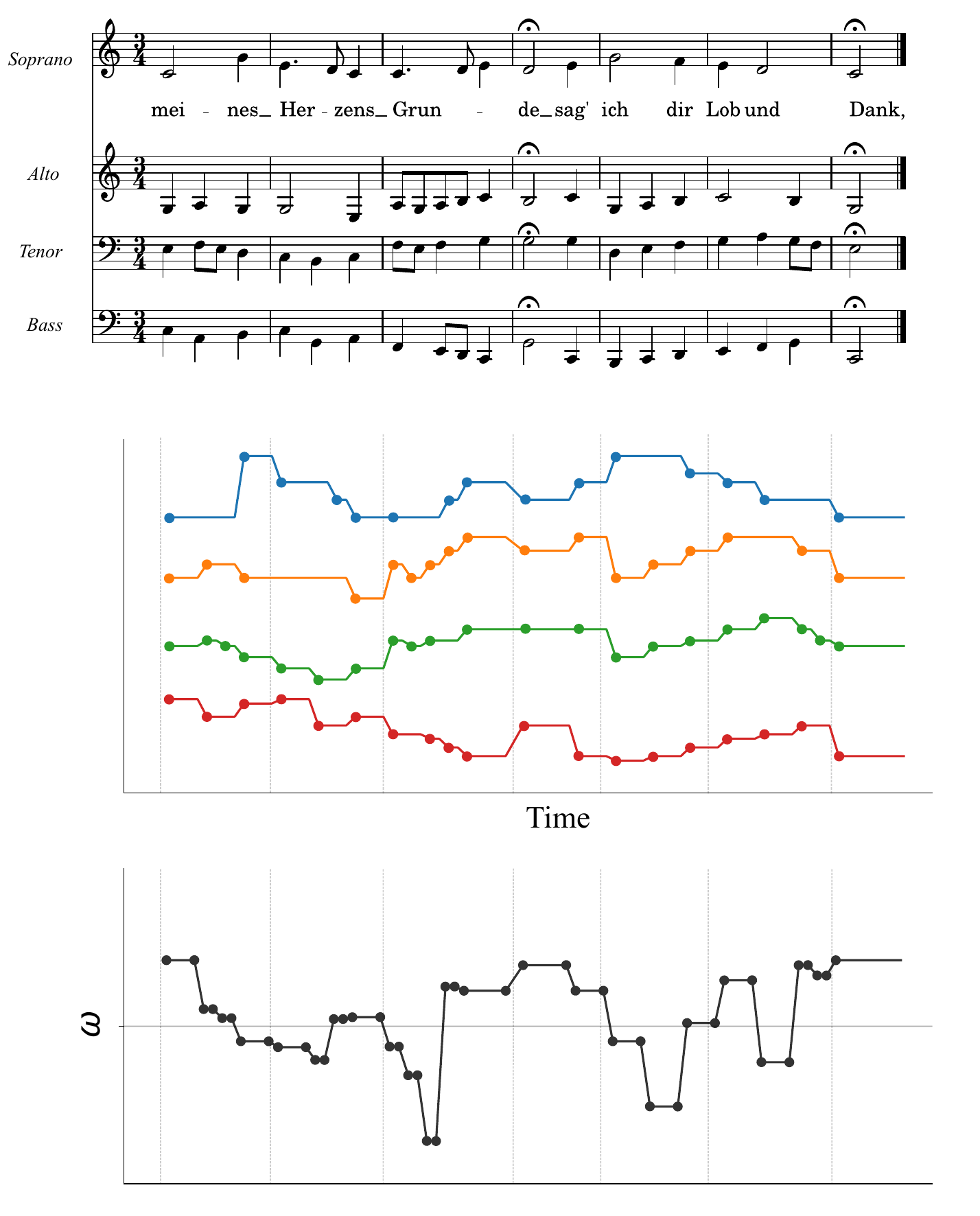}
    \caption{\textbf{Processing pipeline} \textit{Top}: each chorale was selected among those written in major mode and then transposed in C major. \textit{Middle}: each melodic line was transformed into a time series of $13$ possible values. \textit{Bottom}: for each 4-note state the local O-information was calculated using Eq.~\eqref{eq:localO}.}
    \label{fig:Pipeline}
\end{figure}

Our analysis is based on the electronic scores publicly available at
\url{http://kern.ccarh.org}, a website that hosts professionally curated digital scores~\cite{sapp2005online}. Our pre-processing pipeline is the same as in Ref.~\cite{rosas_quantifying_2019}, which we describe here for completeness. The scores were pre-processed
in Python using the \texttt{Music21} package
(\url{http://web.mit.edu/music21}), which allowed us to select only the pieces
written in Major mode. Each chorale was transposed to \textit{C~Major}, and each melodic 
line was transformed into time series of 13 possible values (one for each note plus one
for the silence), using a small rhythmic duration as common time unit.
This resulted in a total of 172 chorales, which gave $\approx 4\times 10^4$ four-note chords.

With this data, the joint distribution of the values for the four-note chords 
was estimated using their empirical frequency~\footnote{\unexpanded{Please note that regularisation methods --- such as Laplace smoothing --- can have significant effects on
the results. We decided not to use such methods, as some chords 
(e.g. C-C$\sharp$-D-D$\sharp$) are not representative of the Baroque repertoire.}}. 
This leads to a probability assigned to each four-note chord, which assess simply
the odds of picking that chord when randomly selecting one out of the whole repertoire.
One can express this probability as the multivariate statistic $p(x_1,x_2,x_3,x_4)$,
with each variable corresponding to the different voices.

Finally, we used $p(x_1,x_2,x_3,x_4)$ to calculate the local O-information $\omega(\bm{x})$ for each chord $\bm{x}$ using Eq.~\eqref{eq:localO}, which determines the dominant
statistical behaviour (in terms of synergy and redundancy) associated with each chord. The overall pipeline starting from the music score and arriving to the local O-information is illustrated in Figure~\ref{fig:Pipeline}.


\subsubsection{Research questions and tools}

We studied the multivariate properties of each of the possible four-note chords of Bach's chorales. 
Our analysis focuses exclusively on harmony and chords, leaving melodic and rhythmic properties to 
future studies. We focus on the question of what harmonic properties of the music tend to 
give rise to synergistic or redundant high-order relationships between the four voices.

Let us denote by $\bm{X}=(X_1,X_2,X_3,X_4)$ the random vector that follow the statistics encapsulated by $p(x_1,x_2,x_3,x_4)$. 
Following standard musical practice, we follow the convention that the variables go from lower to higher range, so that $X_1$ corresponds to the bass and $X_4$ is the soprano. Moreover, we use the shorthand notation $\texttt{CEGE}$ when referring to the chord $(x_1,x_2,x_3,x_4)=(\texttt{C},\texttt{E},\texttt{G},\texttt{E})$.

Note that $\bm{X}$ can adopt $13^4=28561$ possible values, and that $p$ is generally not invariant under changes of ordering between the four voices. Since $X_1,\dots,X_4$ take values among alphabets of cardinality $|\mathcal{X}|=13$, we
do all calculations using logarithms to base
$13$, so that $H(X_k) \leq 1$ for all $k\in\{1,\dots,4\}$ --- we 
call this unit a \emph{mut}, for \emph{musical
bit}. Eq. \eqref{eq:bounds} implies that $-2<\Omega(\bm{X})<2$, and hence most values of $\omega$ are expected to have absolute value less than 2 muts --- which gives a sense of how to interpret the magnitude of local O-information values.

We expected to find a correspondence between tonality and O-information values. In particular, we hypothesise that the principal tonal chords (\texttt{C}, \texttt{F}, and \texttt{G} major) would be associated with redundant behavior, while chords that are farther away from the tonal centre (i.e. involve many sharp or flat alterations) would be related to synergistic events. Additionally, we expect dissonance to be associated with less redundancy, as it involves more complex combinations of notes.

\subsection{Results}~\label{sec:results}

\subsubsection{Analysis of the extreme values of the local O-information}
Out of the $13^4$ possible chords, we found that only $1715$ of them were observed at least once in the chorales, corresponding to only $6\%$ of the possibilities --- reflecting the specificity of the chord choices used in Bach's chorales. A weak correlation is observed between frequency and local O-information: more frequent chords tend to have a higher $\omega$ --- which suggests that more visited chords tend to be made by  more redundant parts (see Figure~\ref{fig:oinfoVsProb}). The most frequently encountered chords are shown in the Appendix (Table~\ref{tab:chords}).
\begin{figure}
    \centering
    \includegraphics[width=0.5\textwidth]{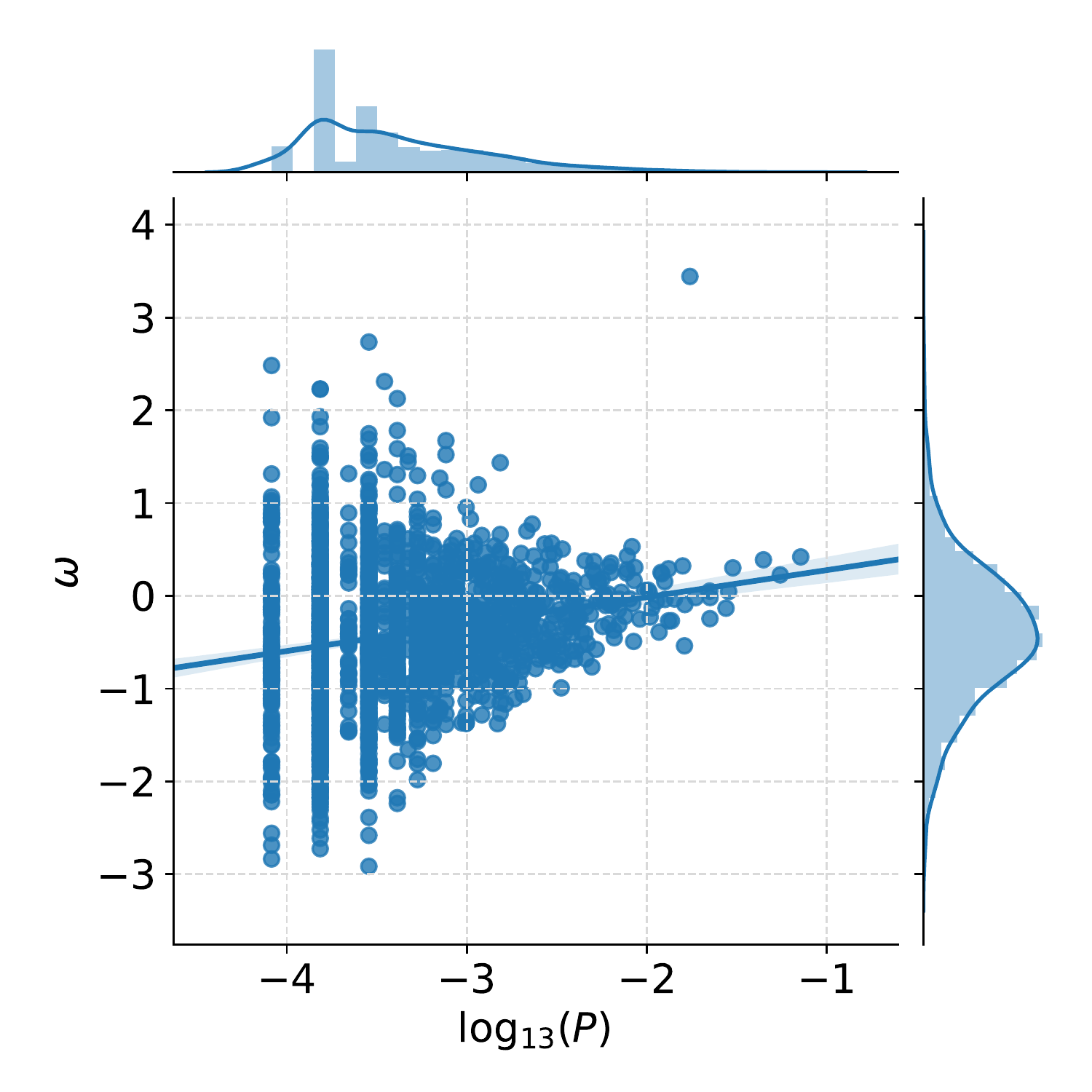}
    \caption{Scatter plot of the logarithm of the probability of each state versus $\omega$. Almost all the states have $\omega$ between $-2$ and $+2$ --- i.e. the bounds for $\Omega$. The outlier in redudancy corresponds to the chord \texttt{RRRR}.}
    \label{fig:oinfoVsProb}
\end{figure}


Some interesting observations can be made by observing the most positive (redundant) and the most negative (synergistic) states in terms of $\omega$, which are presented in Table~\ref{tab:extreme_values}.
On the one hand, the most redundant states tend to contain few alterations (sharp notes, denoted in the table with the symbol $\sharp$) and mostly consonant intervals~\footnote{For a description of consonant and dissonant intervals, please see next section.}. 
In contrast, synergistic chords tend to contain more alterations and dissonant intervals, which in the Western culture are typically associated with harshness and unpleasantness. 
For example, the most synergistic chord contains a major seconds (\texttt{D-E}), while the second most synergistic has one minor second (\texttt{F}$\sharp$\texttt{-G}) and one major second (\texttt{E-}\texttt{F}$\sharp$). 
The ``chord'' with highest local O-information is found to be \texttt{RRRR}, where the redundancy can be interpreted as a consequence of the voices doing the same thing --- not signing.
%
\setlength{\tabcolsep}{10pt}
\begin{table}
    \centering
    \footnotesize\begin{filecontents*}{Table1.csv}
Chord,O-Info,Chord,O-Info,
R R R R,3.443,A E D D,-2.916, 
G D G D,2.736,G B F$\sharp$ E,-2.836, 
F C F C,2.484,B F B B,-2.725, 
A C A C,2.311,A$\sharp$ E E A,-2.688, 
C G C C,2.23,G F$\sharp$ F$\sharp$ A,-2.613, 
E G E G,2.228,G C B A,-2.581, 
C G C G,2.127,F A$\sharp$ G F,-2.559, 
A A E A,1.93,G C C A$\sharp$,-2.522, 
F D G D,1.921,G E C$\sharp$ A,-2.432, 
D D A A,1.824,G G G$\sharp$ C,-2.396, 
G D G G,1.782,R G R E,-2.388, 
D D A D,1.748,G$\sharp$ F G$\sharp$ C,-2.311, 
D F C A,1.688,A$\sharp$ F G$\sharp$ C,-2.276, 
G G D G,1.674,G A F G,-2.245, 
F F C F,1.594,E G A F,-2.238, 
E C E C,1.586,E F$\sharp$ C D,-2.221, 
A C A D,1.544,F$\sharp$ F$\sharp$ C$\sharp$ A,-2.219, 
F F C D,1.532,G F F A$\sharp$,-2.185, 
R R R A,1.522,E A G D,-2.176, 
G F G D,1.512,C$\sharp$ G G B,-2.173, 
F G C D,1.508,G A C$\sharp$ A,-2.173, 
F$\sharp$ A C A,1.506,A D$\sharp$ A D,-2.144, 
B G B G,1.5,E E F$\sharp$ A,-2.143, 
C C G A,1.482,F$\sharp$ B E E,-2.142, 
F A A F,1.46,A G A$\sharp$ F,-2.115, 
G B G B,1.446,F A C$\sharp$ G,-2.114, 
C C G C,1.435,F B G$\sharp$ C,-2.103, 
G G D D,1.361,F$\sharp$ E B D,-2.099, 
G F A D,1.318,G E D A$\sharp$,-2.09, 
C A D E,1.315,C$\sharp$ E F$\sharp$ B,-2.086, 
\end{filecontents*}
\csvautobooktabular[/csv/respect sharp=true, filter={\value{csvrow}<20}, table head=\toprule  \multicolumn{2}{c}{Redundancy} & \multicolumn{2}{c}{Synergy}  \\\cmidrule(r){1-2}  \cmidrule(r){3-4} \bfseries Chord &\bfseries $\omega$ & \bfseries Chord &\bfseries $\omega$ \\\cmidrule(r){1-2}  \cmidrule(r){3-4}]{Table1.csv}
    \caption{\textbf{Chords with the highest (redundance) and lowest (synergy) local O-information.} Letters refers to the standard music nomenclature (plus R is for silence), and the ordering of the voices is Bass-Tenor-Alto-Soprano --- from left to right.}
    \label{tab:extreme_values}        
\end{table}


\subsubsection{The role of intervals}
The results reported in the previous section imply a link between the musical (harmonic) properties of a chord and the type of the statistical interdependencies within it constituent notes. 
In particular, results suggest that synergistic interdependencies may be linked to the presence of dissonances, while redundancy may be related to consonance. A consonant interval occurs when the ratio of the frequencies between two notes is very simple, like $(1\!:\!2)$ for the octave, $(2\!:\!3)$ for the perfect fifth or $(4\!:\!5)$ for the major third. In the Western culture, consonance is typically associated by listeners with pleasantness and acceptability~\cite{lahdelma_cultural_2020}. 
In Westerner music theory, dissonant intervals typically include the major second ($8\!:\!9$) and minor second ($15\!:\!16$), major seventh ($8\!:\!15$) and minor seventh ($9\!:\!16$), and the augmented forth (so-called `tritone' or \textit{diabolus in musica}).

\begin{figure}[htb!]
    \centering
    \includegraphics[width=0.45\textwidth]{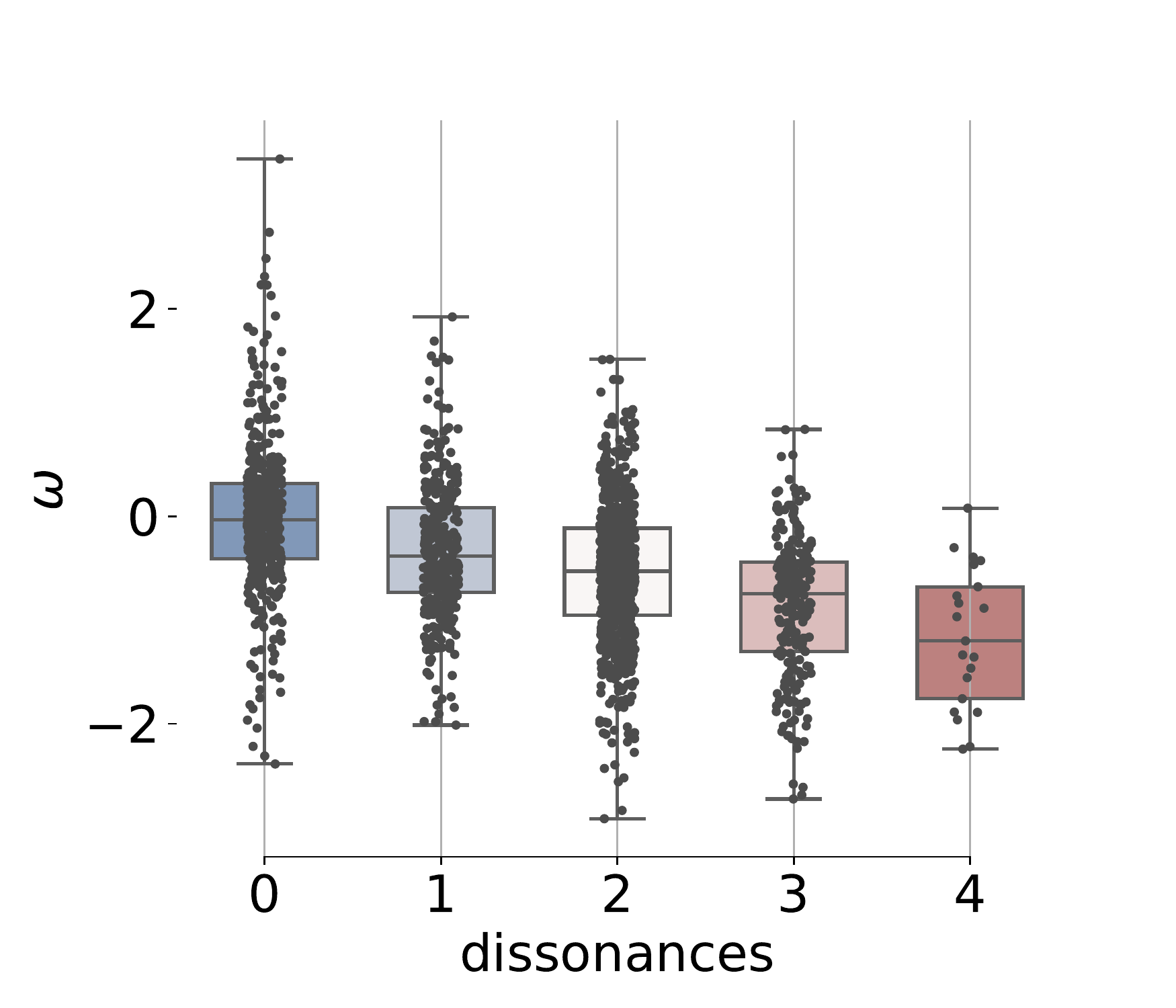}
    \caption{{
    \textbf{Dissonance vs local O-information.}
    \rm Each box represents a state with a different number of dissonant interval. Each category is statistically different from the one with no dissonance (pure consonant intervals).    
    \label{fig:diss}}}
\end{figure}

To further explore the relationship between high-order statistics and harmony, we studied how the local O-information depends on the number of dissonant intervals (either seconds/sevenths or augmented fourths) a chord possesses. The results are depicted in Figure~\ref{fig:diss}. 
Analysis of variance with Bonferroni correction revealed a significant dependency
between number of dissonances and $\omega$ for all number of chords --- except for the contrast between 3 and 4 dissonances, arguably due to the small number of chords with 4 dissonances. The results of all the comparison are shown in Table \ref{tab:diss}.
\begin{table}[!hb]
    \centering
    \footnotesize\begin{filecontents*}{Table2.csv}
0-1,$<0.0001$,0.432
0-2,$<0.0001$,0.738
0-3,$<0.0001$,1.187
0-4,$<0.0001$,1.556
1-2,$<0.0001$,0.317
1-3,$<0.0001$,0.821
1-4,$<0.0001$,1.272
2-3,$<0.0001$,0.488
2-4,$0.0004$,0.924
3-4,$0.5865$,0.447
\end{filecontents*}
\csvautobooktabular[ table head=\toprule  Dissonances  & $p-$value & Cohen's $d$ \\\midrule]{Table2.csv}
    \caption{}
    \label{tab:diss}        
\end{table}
The chords that contain 3 or more dissonant intervals are presented in the Appendix (Table~\ref{tab:dissonances}), most of which exhibits negative values of $\omega$. Remarkably, the only two chords of those with $\omega>0$ can be identified as part of \texttt{G} major with added 7-th, which is the most frequent dissonant chord in classical harmony.

\begin{figure}
    \centering
    \includegraphics[width=0.45\textwidth]{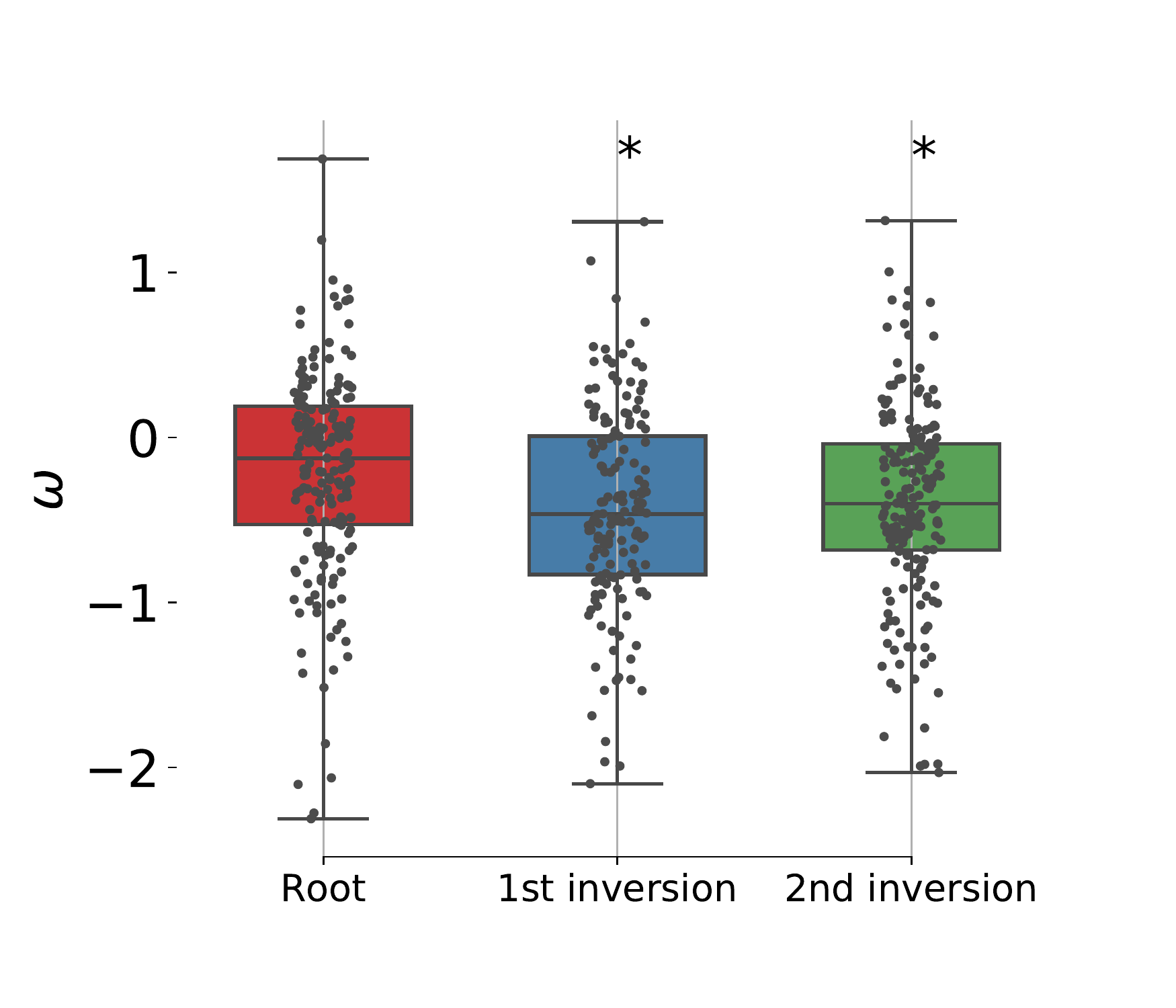}
    \caption{\textbf{The effect of chords inversions on the local O-information.} Root form occurs when the root note of the triad is in the bottom (e.g. \texttt{CEG}), first inversion when the third of the triad is the bass note (e.g. \texttt{EGC}), and second inversion when the fifth is in the bass (e.g. \texttt{GCE}).}
    \label{fig:root_vs_inversion}
\end{figure}

A question that raises from the results shown in Figure~\ref{fig:diss} is why purely consonant chords can be dominantly synergistic, as shown by the variance of the values of $\omega$ for zero dissonance. 
For example, the chord $\omega(\texttt{CEGC})=0.42$ is redundant while $\omega(\texttt{EGCC}) = -0.45$, being both \texttt{C} major chords but having a different pitch in the bass (the fundamental note in the first case, and the third in the second). 
Leveraging music theory, an explanation from 
this comes from the notion of `chord inversion:' 
a triad chord is in \emph{first inversion} if the third (either major or minor) is in the bass, it is in \emph{second inversion} if the fifth is in the bass, and it is in \emph{root position} if the first/fundamental note is in the bass. In Western classical music each inversion tends to be associated with specific sensations -- the first inversion gives a sense of lightness, while the second inversion and root position are typically associated with instability and stability, respectively. 
By considering the values of $\omega$ corresponding to different inversions, t-test shows a tendency towards lower  values of $\omega$ in chords in first (Cohen's $d\simeq 0.36$) and second  (Cohen's $d\simeq 0.31$) inversion when compared to chords in root position, as shown in Figure~\ref{fig:root_vs_inversion}.

\begin{figure*}[!t]
    \centering
    \includegraphics[width=\textwidth, height =0.48\textheight]{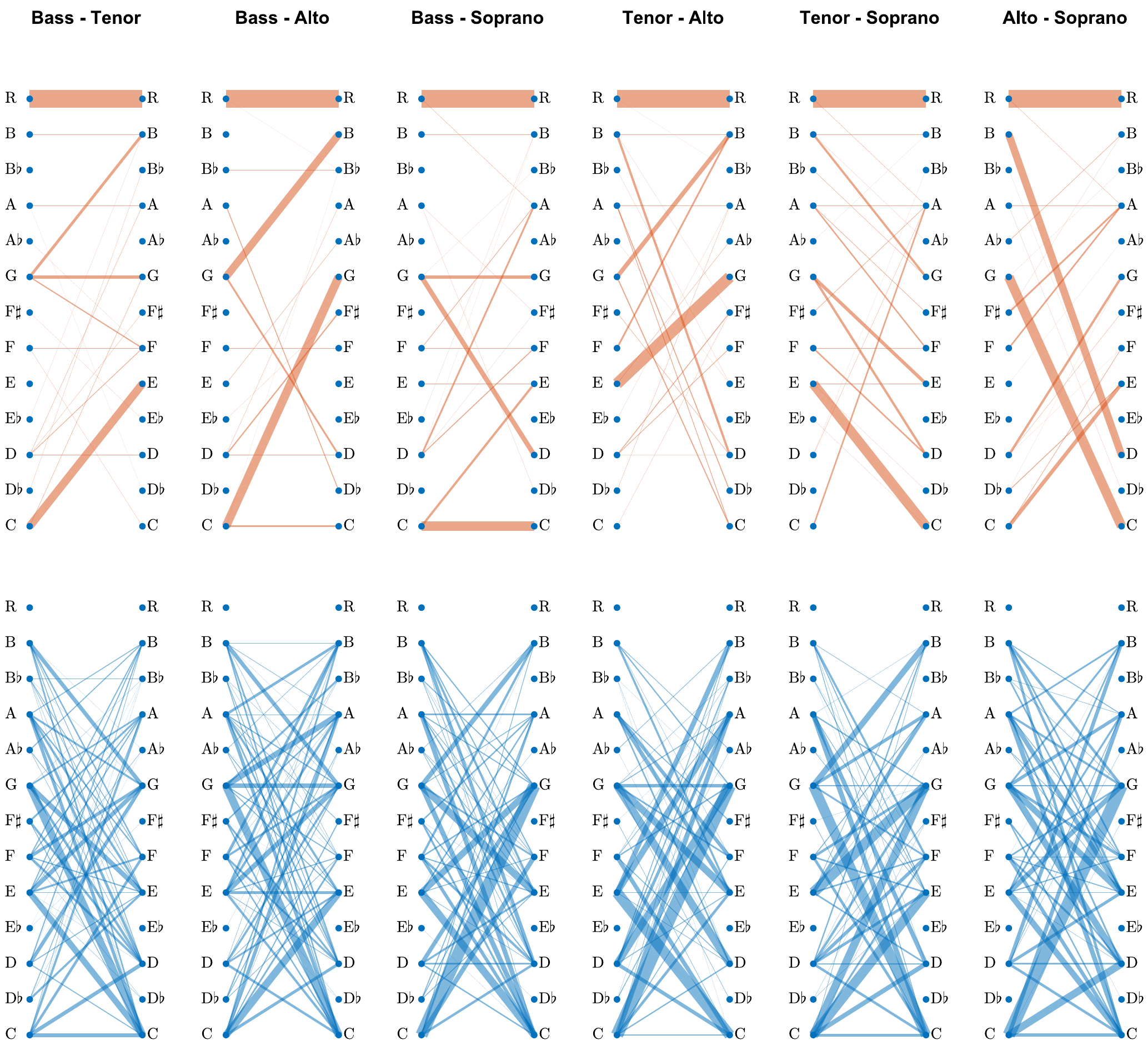}
    \caption{Networks of redundant (orange) and synergy-dominated  (blue) relationships in Bach's chorales. For each voice pair the links were created by averaging the local O-information of all chords that contain that pair of notes.}
    \label{fig:bip_graph}
\end{figure*}

Finally, as a complementary way to study the role of intervals on the O-information, we considered the average value of $\omega$ for given notes at specific voices --- averaging over all possible notes adopted by the other two voices. 
Results are shown in Figure~\ref{fig:bip_graph}. 
It was found that redundancy (i.e. the most positive values of O-information) is `localised' in few intervals, taking place mainly between notes involving the tonic (\texttt{C} major, \texttt{CEG}) or dominant (\texttt{G} major, \texttt{GBD}) chords, or between silences. Also, most redundancy in the bass is associated to the fundamental note of each chord --- either \texttt{C} or \texttt{G}. In contrast, synergy (i.e. the most negative values of O-information) are much more widespread. 
Interestingly, the redundancies between the two extreme voices (soprano and bass) are relatively weak (except between their silence), while synergies between them are not. Please note that the extreme voices tend to carry an important role in Bach chorales --- the soprano carrying out the main melody, and the bass leading the harmony. 
%


\subsubsection{Harmonic depth}


While the previous section focused on the role of single intervals, now our analyses focuses on harmonic considerations. 
Harmony is organised around a \emph{tonality} (also called `key'), which plays the role of centre of gravity around which music discourse revolves. 
The axial pitch of a tonality is called the \emph{root}, which in turn gives name to the tonality --- e.g. \texttt{C} is the root of the tonality of \texttt{C} major. 
In classical Western music there are 12 different major tonalities, one for each of each pitch. 
Also, each major tonality has an associated minor tonality, which is located a minor third below (e.g. \texttt{C} major is associated with \texttt{A} minor). 
Each of these 12 major tonalities are made of 7 distinct pitches, and are naturally ordered by a notion of proximity depending on how many pitches do they have in common. This gives raises to the \emph{circle of fifths}: major keys separated by a fifth have only one note different. For example, \texttt{C} and \texttt{G} major are only distinguished by the note \texttt{F}, which is sharp for the latter but natural for the former. 

\begin{figure}[htb!]
    \centering
    \includegraphics[width=0.45\textwidth]{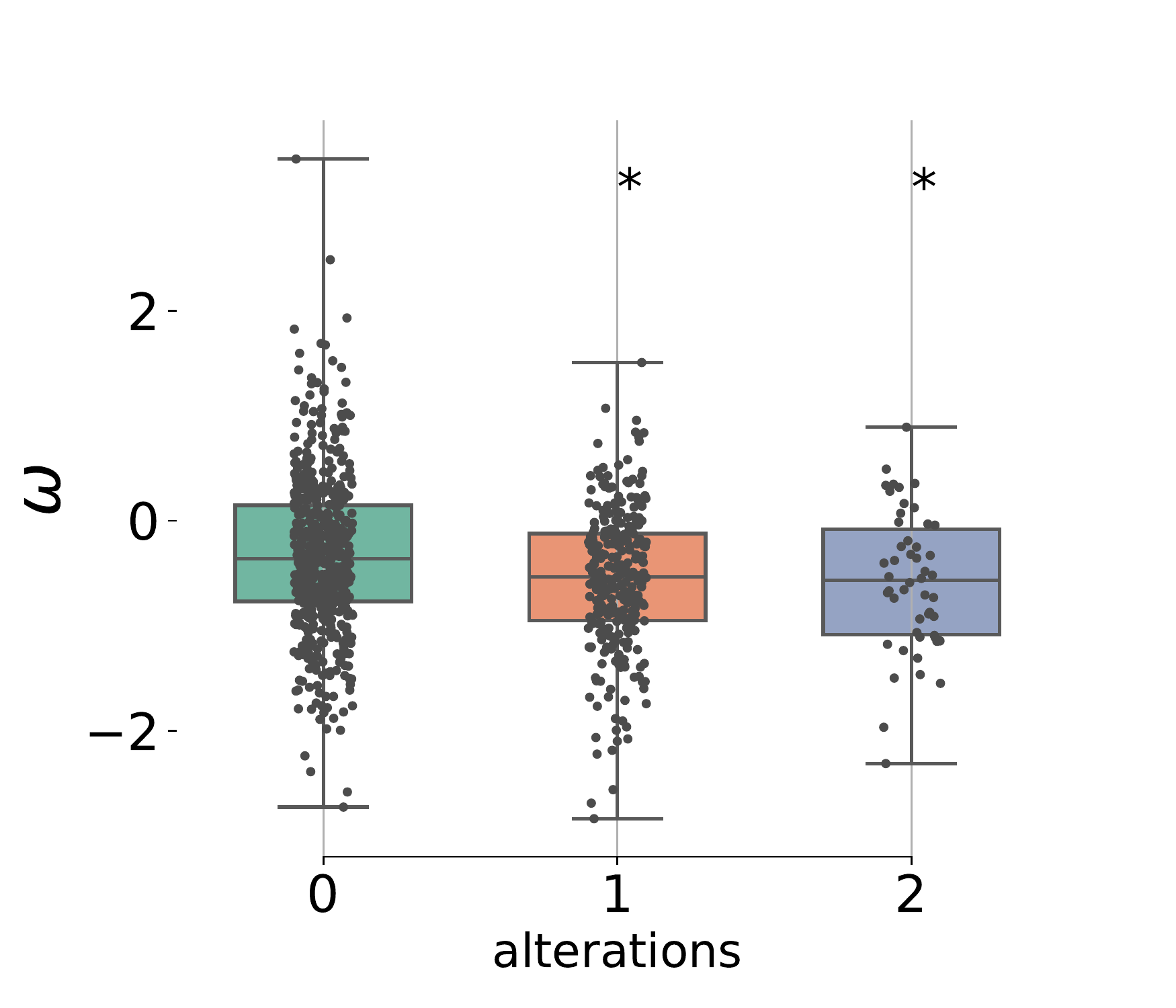}
    \caption{\textbf{The local O-information is related to the number of alterations (sharps and flat notes) inside each state.} States with 1 and 2 alterations have $\omega$ significantly lower ($p<0.05$) than states with no alterations. }
    \label{fig:sharps}
\end{figure}

A simple way to explore the impact of harmony on the high-order statistics is by analysing the dependency between $\omega$ and the number of alterations (sharps or flats) that a chord has. In effect, please recall that all the chorales analysed are in major mode, and have been shifted to \texttt{C} (see Section~\ref{sec:pipeline}). 
Moreover, chords belonging to \texttt{C} major have no alterations, while chords for more distant tonalities have progressively more alterations --- either sharps if going up the cycle of fifths, or flats otherwise. Therefore, we ran statistical analyses (t-test corrected for multiple comparisons) on the effect of the number of alterations on $\omega$, whose results are shown in Figure~\ref{fig:sharps}. Results revealed significantly decreases of $\omega$ ($p<0.01$) for states with one (Cohen's $d\simeq 0.41$) or two (Cohen's $d\simeq 0.44$) alterations with respect to chords without any alterations. 
\begin{figure}[htb!]
    \centering
\includegraphics[width=0.45\textwidth]{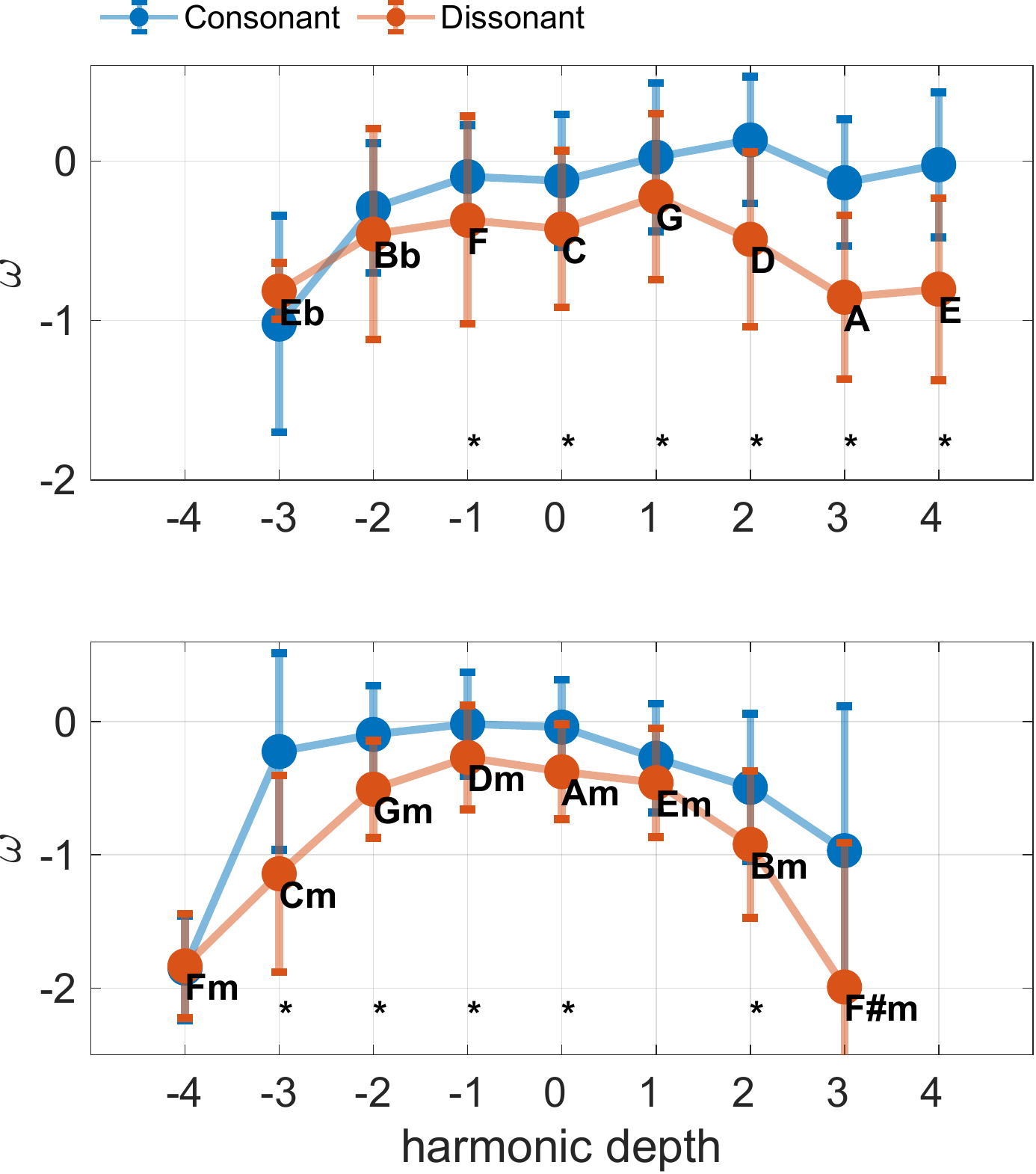}
    \caption{\textbf{The effect of harmonic distance.} \rm The figure shows the mean values (and confidence intervals) of $\omega$ for musical states in every different major (top) and minor (bottom) chord. The blue line indicates purely consonant triads, while the orange line corresponds to chords with one or more dissonances.}
    \label{fig:harmDepth}
\end{figure}

In order to go deeper into the impact of harmony on the O-information we introduced the notion of \emph{harmonic depth}, which corresponds to 
the smallest number of steps (going clockwise or anti-clockwise) in the circle of fifths are required to go from $C$ major to a given tonality --- or from $A$ minor, in case of minor tonality. For example, the chord $D$ major has an harmonic depth of $+2$, while the chord \texttt{F} minor (denoted as \texttt{F}m) has an harmonic depth of $+4$. 

We are interested to study the relationship between $\omega$ and 
harmonic depth. 
For this purpose, we consider the values of $\omega$ for triads that belong to a specific tonality, regardless to the arrangement of notes between the voices 
For example, determining a chord is \texttt{C} just accounts for triads containing only the pitches 
\texttt{C,E,G}, regardless how they are arranges among the voices.
Results are shown in Figure~\ref{fig:harmDepth}, and show that the value of $\omega$ decreases as soon as the tonality moves away from \texttt{C} major, being this difference more pronounced in minor chords. 
This suggest that synergy may also be associated with more complex harmonic explorations involving more harmonically distant chords.
%


\subsubsection{Lyric analysis}

As a final step in our analysis, we investigated the relationship between the values of $\omega$ due to the chords and the corresponding word that is sang as part of the lyrics. 
For this purpose, we consider the different values of $\omega$ that correspond to each time a given word is sang. In cases of melismas (i.e. when many notes are sang under the same syllable), the values of the whole progression were averaged and counted as one realisation of the word.

\begin{figure}[h!]
    \centering
    \includegraphics[width=0.45\textwidth]{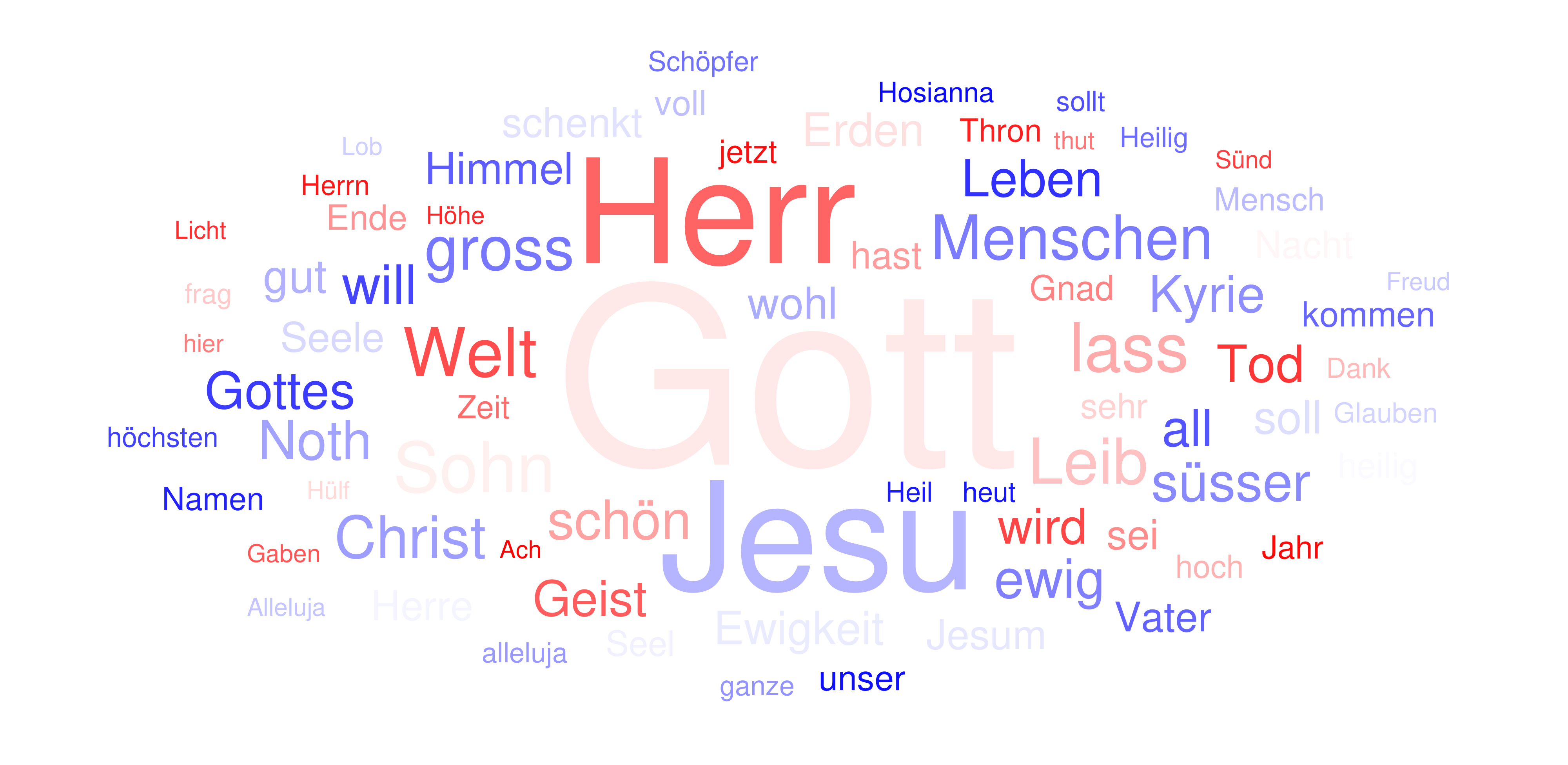}
    \caption{\textbf{Word cloud of Bach's chorales lyrics.} Most common words found in text, with their size representing their frequency and their colour their local O-information sign (plus or minus).}
    \label{fig:wordcloud}
\end{figure}

As a first analysis, we calculated a word cloud where words associated with negative or positive values of $\omega$ are represented in red and blue, respectively.
As shown in Figure ~\ref{fig:wordcloud}, many of the most common words (like \textit{Gott}, \textit{Herr}, \textit{Sohn}) are redundant, with the exception of \textit{Jesu} that is synergistic. 
As the majority of the chords explored by Bach are synergistic (the average O-information is negative, see Figure~\ref{fig:oinfoVsProb}), this prevalence of redundant words is highly non-trivial. 

\begin{figure}[h!]
    \centering
    \includegraphics[width=0.45\textwidth]{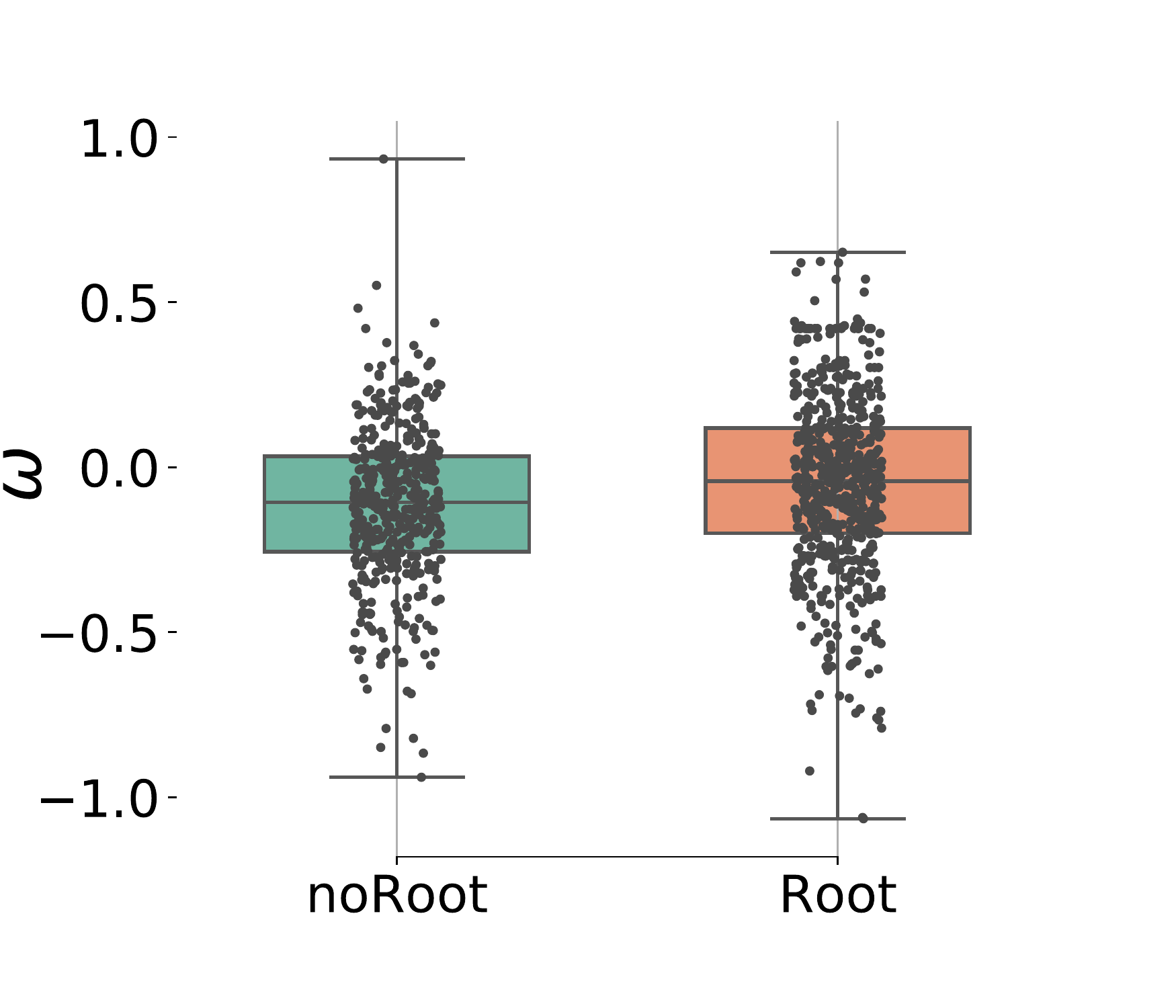}
    \caption{\textbf{Root vs non root words.} Each box shows the distribution of local O-information $\omega$ for each category. A statistical comparison between the two populations has been carried out with a two sample $t$-test, which rejected the null hypothesis with $p<0.01$; the effect size can be expressed as Cohen's $d\simeq 0.26$.}
    \label{fig:rootWords}
\end{figure}

Another insight that can be drawn from the word cloud is that words that are not the subject of the phrase seems to be more synergistic. 
To verify this, we evaluated the effect on $\omega$ of words being in root form (nominative case) with respect to all others~\footnote{German language has four cases: nominative (subject), accusative (direct object), dative (indirect object), and genitive (possessive).}. 
Results confirmed our conjecture, showing that words in root form have a tendency towards higher values of $\omega$ (see Figure~\ref{fig:rootWords}). As an speculation, this may be interpreted by noting that root form words correspond to the most important part of the sentence, and hence a redundant underlying harmony might contribute to a easier comprehension. The most frequent cases of words for which we found both the root and the non root word are shown in Table \ref{tab:commonWords}.
\begin{table}[!hb]
    \centering
    \footnotesize\begin{filecontents*}{Table3.csv}
Root,oiRoot,NoRoot,oiNoRoot
Gott,0.008,Gottes,-0.233
,,Gotte,-0.002
,,Gotts,-0.256
Herr,0.096,Herre,-0.007
,,Herren,-0.152
Christ,-0.127,Christen,-0.442
,,Christe,-0.092
,,Christenheit,-0.342
ewig,-0.159,Ewigkeit,-0.024
,,ewiglich,-0.264
Geist,0.098,Geistern,-0.866
soll,-0.042,sollen,-0.111
,,sollst,-0.444
,,solls,-0.009
Himmel,-0.188,Himmels,-0.314
Seel,-0.02,Seele,-0.044
Gnad,0.084,Gnaden,0.052
,,Gnade,0.066
Seel,-0.02,Seelen,0.083
voll,-0.068,voller,-0.458
Mensch,-0.07,Menschen,-0.16
hoch,0.059,höchsten,-0.224
,,Höchsten,-0.158
,,höchster,-0.201
Ehr,-0.277,Ehren,-0.222
Licht,0.137,Lichtes,-0.288
Ehr,-0.277,Ehre,-0.107
ohn,-0.214,ohne,-0.336
Trost,-0.286,Tröster,-0.236
End,0.106,Ende,0.074
Herz,0.175,Herzen,0.132
,,Herzens,-0.149
,,Herze,0.32
\end{filecontents*}
\csvautobooktabular[/csv/respect sharp=true, table head=\toprule\bfseries Root  &\bfseries $\omega$ &\bfseries No Root &\bfseries $\omega$ \\\midrule]{Table3.csv}
    \caption{Most common words and their mean local O-information value.}
    \label{tab:commonWords}        
\end{table}

\section{Conclusions}

This paper introduces a new framework to study the high-order interdependencies observed in complex multivariate systems, which is capable of disentangling their effects on individual patterns of activity. 
The approach is centred on the local O-information, a measure that quantifies the balance between redundancy and synergy at each pattern. 
Because of its information-theoretic nature, this measure is widely applicable --- being suitable to assess systems with categorical, discrete, and continuous variables.

The capabilities of the proposed framework were showcased in an analysis of the scores of the chorales of J.S. Bach, which illuminated the high-order relationships that exist between the different voices. In particular, our results synergy-dominated interdependencies tend to be associated with complex musical elements, including dissonances, chord inversions, and harmonic distance from the tonal centre. Taken together, our findings provide converging evidence about the relationship between statistical synergy and the complexity of the musical discourse.

These findings have interesting parallels with recent studies on the human brain, which are revealing a close relationship between synergistic interdependencies in neural activity and high cognitive functions. 
Historically, the notion of synergistic information was originated in theoretical neuroscience as an effort to characterise aspects of complex neural activity~\cite{tononi_measure_1994,tononi_complexity_1998,gat_synergy_1999,schneidman_synergy_2003,latham_synergy_2005,ganmor_sparse_2011}. 
Moreover, recent empirical work has shown that synergy characterises the interactions between multimodal, association brain areas that have undergone more evolutionary expansion~\cite{luppi_synergistic_2020}, and is also associated with changes in brain function due to anaesthesia and disorders of consciousness~\cite{luppi_synergistic_2020-1}. 
Under the light of these findings, the results presented in this work let us speculate that elaborated musical discourse may have key similarities with the type of neural activity that underpins high brain functions. 
One possible commonality could be presence of emergent phenomena, which have been recently characterised formally in terms of statistical synergy~\cite{rosas_reconciling_2020}. 
This would not be the first time music is shown to share some of the hallmark properties of complex systems; in effect, properties of musical discourse has been show to be related to non-linear fluctuations and self-organised criticallity~\cite{telesca_revealing_2011,gonzalez-espinoza_multiple_2017}, and also to entropy production and irreversibility~\cite{gonzalez-espinoza_arrow_2020}.

Pointwise information measures, initially proposed in Ref.~\cite{lizier_local_2008} w.r.t. the local transfer entropy, are a promising set of techniques that can provide a detailed description of information transfer mechanisms in complex systems.  Recently this paradigm has been applied to implement the local Granger causality \cite{stramaglia_local_2021}, which has shown interesting results on physiological and neural data. The fine descriptions allowed by the formalism introduced in this paper bring a new perspective over high-order interdependencies, which complement existent pointwise information decomposition approaches (e.g. Ref.~\cite{finn_pointwise_2018-1}) by being applicable to larger systems --- hence greatly extending their domain of practical applicability. The extension of these ideas to dynamical information decomposition, such as the integrated information decomposition framework~\cite{mediano_beyond_2019}, constitutes a promising direction for future research.

\section*{Acknowledgements}

The authors thank Pablo Padilla and Alejandro Reyes for insightful discussions. 
T.S. wants to thank M. Paolo Daniele for having transmitted to him a bit of his passion for music. 
F.R. was supported by the Ad Astra Chandaria foundation. S.S. was supported by MIUR project PRIN 2017WZFTZP ``Stochastic forecasting in complex~systems''.

\section*{Code availability}
Code for the analysis reported here is publicly available at \url{https://github.com/tomscag/local_O_information}

\clearpage
\newpage

\appendix

\section{Tables}\label{sec:appendix_tables}
\begin{table}[!hb]
    \centering
    \footnotesize\begin{filecontents*}{Table4.csv}
Chord,Oinfo,Occurrence
C E G C,0.4201,1881
C G C E,0.226,1405
G G B D,0.3887,1106
G B D G,0.3019,715
G B G D,0.0464,674
C C G E,-0.1328,649
G D B G,-0.0182,517
C G E C,-0.2457,516
A E A C,0.0517,514
G D G B,-0.0189,421
R R R R,3.4431,388
E G C G,-0.0938,361
C E C G,-0.5384,359
G F B D,0.3229,350
B D G D,-0.0342,311
E C G C,-0.2682,298
E G C E,0.2924,286
C C E G,-0.2694,286
B G D G,-0.0387,272
D A D F,0.1538,271
A A C F,0.2383,260
F C F A,0.2522,257
G G C D,-0.3909,250
F F C A,-0.058,241
F F A C,-0.1293,226
F A C D,-0.2281,220
D F A D,0.0358,217
A A C E,0.0617,214
A C E C,0.0578,203
A A E C,-0.2498,190
C E G E,0.307,177
F A C F,0.1666,175
G G D B,-0.491,174
B G D F,-0.0853,172
D D F# A,0.5305,170
\end{filecontents*}
\csvautobooktabular[/csv/respect sharp=true, table head=\toprule\bfseries Chord  &\bfseries $\omega$ &\bfseries Frequency \\\midrule]{Table4.csv}
    \caption{\textbf{Most common chords.}}
    \label{tab:chords}        
\end{table}

\begin{table}[!hb]
    \centering
    \footnotesize\csvautobooktabular[/csv/respect sharp=true, table head=\toprule  \multicolumn{2}{c}{Redundancy} & \multicolumn{2}{c}{Synergy}  \\\cmidrule(r){1-2}  \cmidrule(r){3-4} \bfseries Chord &\bfseries $\omega$ & \bfseries Chord &\bfseries $\omega$ \\\cmidrule(r){1-2}  \cmidrule(r){3-4}]{Table1.csv}
    \caption{\textbf{Table shows the first states with the highest (redundance) and lowest (synergy) $\omega$ values. The letters refers to the standard nomenclature of notes in music (R is for rest) and the order of the voices is Bass-Tenor-Alto-Soprano, from left to right.}}
    \label{tab:extreme_values_all}        
\end{table}

\begin{table*}
\centering
\makebox[0pt][c]{\parbox{1.0\textwidth}{%
    \begin{minipage}[t]{0.40\hsize}\centering
            \small\begin{filecontents*}{Table5.csv}
Chord,O-Info,Occurrence 
C E G C,0.42,1881
C G C E,0.226,1405
C C G E,-0.133,649
C G E C,-0.246,516
E G C G,-0.094,361
C E C G,-0.538,359
E C G C,-0.268,298
E G C E,0.292,286
C C E G,-0.269,286
C E G E,0.307,177
E G C C,-0.452,132
E C G G,-0.302,122
E C G E,0.015,113
E G G C,-0.577,102
E E G C,0.37,99
G G C E,0.149,90
E E C G,0.002,78
E G E C,-0.045,74
E C E G,0.093,71
E C C G,-0.467,52
C C E C,0.442,52
C C E E,0.033,44
C E C E,0.774,41
C G E G,-0.106,37
G G E C,-0.812,35
G E C E,-0.977,26
C C G C,1.435,26
E E G G,-0.16,25
C G E E,0.538,24
G E C C,-1.132,22
\end{filecontents*}
\csvautobooktabular[/csv/respect sharp=true, table head=\toprule\bfseries Chord  &\bfseries $\omega$ &\bfseries Occurrence \\\midrule]{Table5.csv}
            \caption{\textbf{Most common tonal chords.  }}
            \label{tab:tonalchords} 
    \end{minipage}
    \hfill
    \begin{minipage}[t]{0.40\hsize}\centering
            \small\begin{filecontents*}{Table6.csv}
Chord, O Info,Occurrence
G G B F,-0.2391,81
D D F$\sharp$ C,-0.4532,28
G F G B,-0.8748,27
C A$\sharp$ C E,-0.1148,27
D D C F$\sharp$,-0.5139,26
G G F B,-0.8308,25
G F B G,0.226,25
G B G F,-0.7421,24
B A C F,-0.455,22
E F B D,-0.3088,22
D F$\sharp$ D C,-0.713,18
D C D F$\sharp$,-0.7643,17
E D G$\sharp$ E,-0.262,17
B C A F,-0.7095,16
G B F G,0.2502,16
E F B G,-0.4072,14
C C E A$\sharp$,-0.2633,14
C E D E,-0.7414,13
G F B C,-0.8746,12
D F C E,-0.5583,12
C D F$\sharp$ D,-1.0235,10
E E D G$\sharp$,-0.7605,10
D C F$\sharp$ D,-0.1526,10
F B F A,-0.1821,9
B C F C,-1.8124,8
G F A$\sharp$ A,-1.5336,8
E D E G$\sharp$,-0.7108,8
A E A$\sharp$ C,-0.6605,8
F G C$\sharp$ E,-0.5848,8
B C F A,-0.5059,8
\end{filecontents*}
\csvautobooktabular[/csv/respect sharp=true, table head=\toprule\bfseries Chord  &\bfseries   $\omega$ &\bfseries Occurrence \\\midrule]{Table6.csv}
            \caption{\textbf{Most common dissonant chords (only chords with 3 or more dissonant intervals are shown).}}
            \label{tab:dissonances}  
    \end{minipage}
}}
\end{table*}

\clearpage
\bibliographystyle{unsrt}
\bibliography{biblio}

\end{document}